%% file: paper.tex
\documentclass[sigconf]{acmart}
\usepackage[font=small]{caption}
\usepackage[labelformat=simple]{subcaption}
\usepackage{graphicx}
\usepackage[ruled,vlined,linesnumbered]{algorithm2e}
\usepackage{algorithmic}
\usepackage[mathscr]{eucal}
\usepackage{color}
\usepackage{url}
\usepackage{mathtools}
\usepackage{xcolor}
\usepackage{multirow}
\usepackage{soul}
\usepackage{colortbl}

\allowdisplaybreaks

\newcommand{\method}{\textsc{ATR}\xspace}
\newcommand{\ft}{\textsc{ATR-2FT}\xspace}
\newcommand{\icl}{\textsc{ATR-ICL}\xspace}
\newcommand{\opt}{\textsc{OPT}\xspace}
\newcommand{\pointer}{\textsc{POINTER}\xspace}
\newcommand{\hmf}{\textsc{HybridMF}\xspace}
\newcommand{\unisrec}{\textsc{UniSRec}\xspace}
\newcommand{\fdsa}{\textsc{FDSA}\xspace}
\newcommand{\argkdd}{\textsc{ARG}\xspace}

\newcommand{\amr}{\textsc{AMR}\xspace}
\newcommand{\bae}{\textsc{BAE}\xspace}
\newcommand{\checklist}{\textsc{CheckList}\xspace}
\newcommand{\atwot}{\textsc{A2T}\xspace}
\newcommand{\gpt}{\textsc{GPT-2}\xspace}

\newcommand{\gptfour}{\textsc{GPT-4}\xspace}
\newcommand{\llama}{\textsc{Llama-2}\xspace}
\newcommand{\falcon}{\textsc{Falcon-40B}\xspace}

\copyrightyear{2024}
\acmYear{2024}
\setcopyright{rightsretained}
\acmConference[CIKM '24]{Proceedings of the 33rd ACM International Conference on Information and Knowledge Management}{October 21--25, 2024}{Boise, ID, USA}
\acmBooktitle{Proceedings of the 33rd ACM International Conference on Information and Knowledge Management (CIKM '24), October 21--25, 2024, Boise, ID, USA}
\acmDOI{10.1145/3627673.3679592}
\acmISBN{979-8-4007-0436-9/24/10}
\ccsdesc[500]{Information systems~Recommender systems}

\keywords{Text Rewriting Attack, Text-aware Recommender Systems, Model Robustness,  Large Language Models, 
Automated Text Generation}

\begin{document}

\title{Adversarial Text Rewriting for Text-aware Recommender Systems}

\author{Sejoon Oh}
\email{soh337@gatech.edu} 
\affiliation{%
  \institution{Georgia Institute of Technology}
  \city{Atlanta}
    \country{United States}
}

\author{Gaurav Verma}
\email{gverma@gatech.edu} 
\affiliation{%
  \institution{Georgia Institute of Technology}
  \city{Atlanta}
    \country{United States}
}

\author{Srijan Kumar}
\email{srijan@gatech.edu}
\affiliation{%
  \institution{Georgia Institute of Technology}
  \city{Atlanta}
    \country{United States}
}

	\begin{abstract}
		\label{sec:abstract}
		\input{000abstract}
	\end{abstract}

\maketitle

	\section{\textbf{Introduction}}
	\label{sec:intro}
	\input{010introduction}

	\section{Related Work}
	\label{sec:related_work}
	\input{020related_work}

	\section{Problem Formulation}
    \label{sec:preliminaries}
	\input{030preliminaries}

	\section{Proposed Methods}
	\label{sec:proposed_method}
	\input{040proposed_method}

	\section{Experiments: Evaluation of \method}
	\label{sec:experiment}
	\input{050experiment}

	\section{Discussion and Conclusion}
	\label{sec:conclusion}
	\input{070conclusion}

\section*{Acknowledgments}
	{
	This research is supported in part by Georgia Institute of Technology, IDEaS, and Microsoft Azure. S.O. was partly supported by ML@GT, Twitch, and Kwanjeong fellowships. We thank the reviewers for their feedback.
	}	

        \appendix
	\section{Appendix}
	\label{sec:appendix}
	\input{080appendix}

\bibliographystyle{ACM-Reference-Format}
\bibliography{reference}

\end{document}

%% file: 000abstract.tex
Text-aware recommender systems incorporate rich textual features, such as titles and descriptions, to generate item recommendations for users.  
The use of textual features helps mitigate cold-start problems, and thus, such recommender systems have attracted increased attention.
However, we argue that the dependency on item descriptions makes the recommender system vulnerable to manipulation by adversarial sellers on e-commerce platforms. 
In this paper, we explore the possibility of such manipulation by proposing a new text rewriting framework to attack text-aware recommender systems. We show that the rewriting attack can be exploited by sellers to unfairly uprank their products, even though the adversarially rewritten descriptions are perceived as realistic by human evaluators. Methodologically, we investigate two different variations to carry out  text rewriting attacks: (1) two-phase fine-tuning for greater attack performance, and (2) in-context learning for higher text rewriting quality. 
Experiments spanning 3 different datasets and 4 existing approaches demonstrate that recommender systems exhibit vulnerability against the proposed text rewriting attack. 
Our work adds to the existing literature around the robustness of recommender systems, while highlighting a new dimension of vulnerability in the age of large-scale automated text generation.

%% file: 010introduction.tex
Recent recommender systems have increasingly leveraged textual product descriptions as an important input feature~\cite{kim2016convolutional, zhou2019content,shalaby2022m2trec}. 
Such text-aware recommender systems alleviate cold-start problems~\cite{shalaby2022m2trec, chuang2020tpr} and overcome training challenges like underfitting due to data sparsity by incorporating rich textual features~\cite{truong2021multi, salah2020cornac}. Due to their effectiveness, text-aware recommender systems have been employed in many applications including e-commerce~\cite{shalaby2022m2trec}, housing~\cite{shen2020text}, and social platforms~\cite{cheng2022engage}. 

\begin{figure}[t!]
    \centering
    \includegraphics[width=0.9\linewidth]{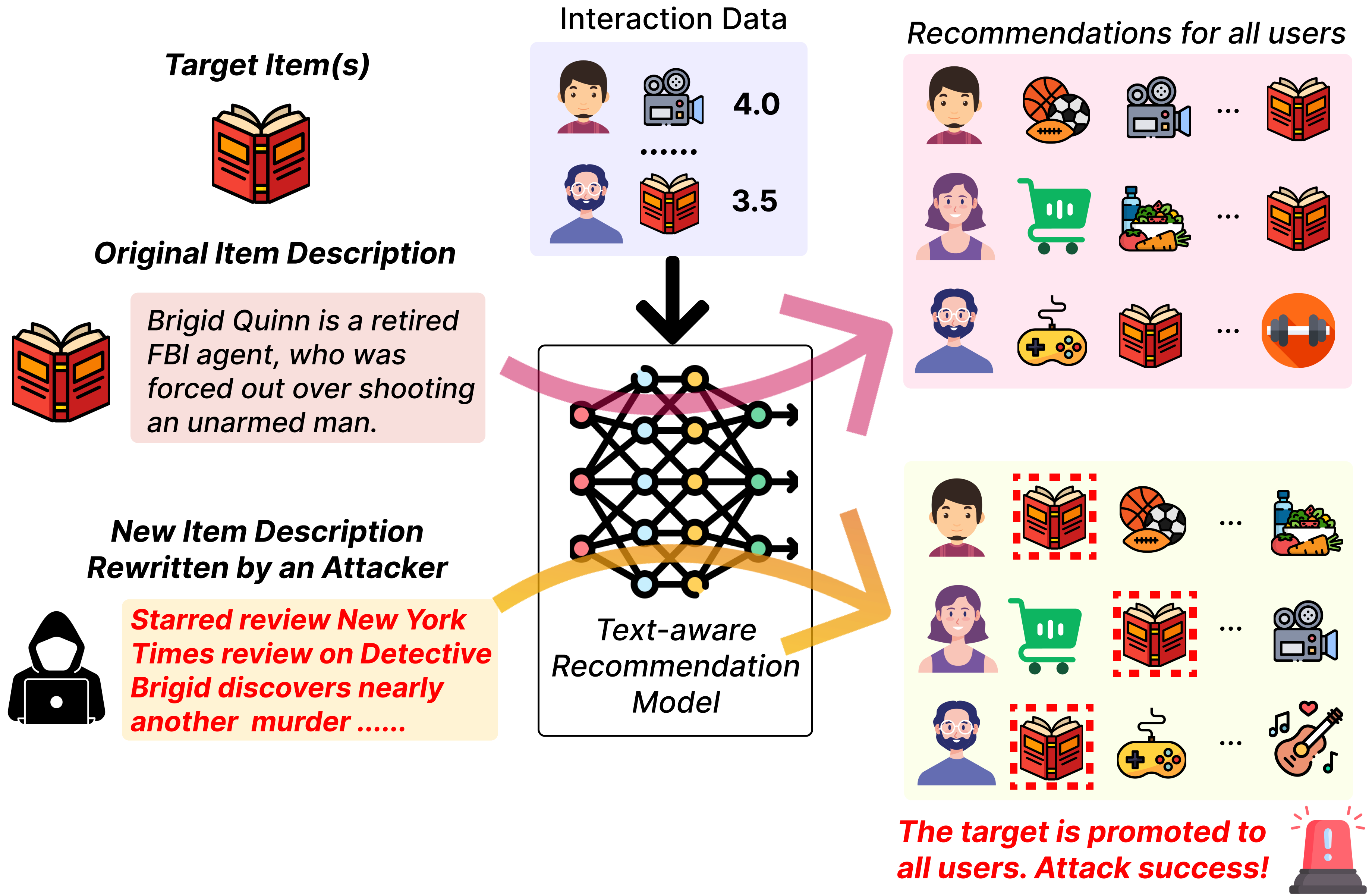}
    \caption{Our work investigates the vulnerabilities of text-aware recommender systems against adversarial product description rewriting that cause increase in ranks of the targeted items across all users.
    }
    \label{fig:intro_figure}
\end{figure}

While text descriptions of items offer the advantages mentioned earlier, our work shows that adversarial sellers can manipulate these descriptions, exposing a new aspect of vulnerability in these recommender systems. 
Consider the scenario where a seller wants to uprank their item(s) to all the users. 
The seller can rewrite the textual description of the item such that the resulting ranking of the item(s) increases across all the users. We consider this as a \textit{text rewriting attack}. 
Figure~\ref{fig:intro_figure} depicts the attack scenario.
Such sellers could be motivated to manipulate the descriptions as  higher ranking translates to increased views, purchases, and revenue without enhancing the item's actual quality~\cite{jannach2019measuring, lappas2016impact, wang2012bonus}. These attacks can be detrimental to all stakeholders (e.g., customers, vendors, platform) in the ecosystem, since the attackers can corrupt the recommendation results with their target products and lower the overall trustworthiness of recommendations.

\begin{table*}[t!]
\small
	\centering
	\caption{Comparison of our proposed adversarial text rewriting framework \method to existing text attack/generation methods. Our proposed approach is the only one that performs targeted attacks and does in-context learning while also incorporates the rank promotion objective.
	}
	\begin{tabular}{c|c|cccccc}
		\toprule
  & \textbf{\method}
        &   \atwot~\cite{yoo2021towards}
		&   \bae~\cite{garg2020bae} 
		&  \checklist~\cite{ribeiro2020beyond}   
		&  \argkdd~\cite{chiang2023shilling} &  
       \gptfour~\cite{openai2023gpt4} &  
       \llama~\cite{touvron2023llama}
		 \\
		\midrule
		Item Description Rewriting Capability & \checkmark &  \checkmark  & \checkmark  & \checkmark  & &  \checkmark & \checkmark   \\
		Perform Targeted Attacks & \checkmark &   &     & \checkmark & \checkmark & &       \\
    	In-context Learning Capability & \checkmark &   &  & &  &  \checkmark  & \checkmark  \\
		Involve a Rank Promotion Objective & \checkmark  &   &  &  & \checkmark &  &  \\
		\bottomrule
	\end{tabular}
	\label{tab:comparators}
\end{table*}

Text rewriting attacks are highly plausible as sellers often have the agency to edit the descriptions of the items they sell. 
This attack is more practical than shilling and injection attacks in recommender systems that require injections of fake users, items, interactions, or reviews~\cite{zhang2021data,zhang2021reverse, song2020poisonrec, surma2020hacking, khalid2018security, Manipulating_ML, juuti2018stay}, because generating such fake entities that look realistic can be computationally heavy and often detected by a fraud detection algorithm. With recent advances in large language models (LLMs)~\cite{touvron2023llama, openai2023gpt4, falcon40b}, it is pertinent to investigate how automated text generation can be leveraged to manipulate product descriptions, so that the rank of the items targeted by malicious sellers is boosted across all users~\cite{deldjoo2021survey, di2020taamr, yue2021black, zhang2020practical}.

To quantify the vulnerabilities of text-aware recommender systems, we investigate approaches to generate adversarial settings. There are three challenges for adversarial text rewriting of target items, given a text-aware recommender system. 
First, generating the rewritten text has two competing objectives: \textit{(a)} promoting the target items' ranks versus \textit{(b)} generating text that is fluent, realistic, and preserves the semantic meaning present in the original description. 
For instance, the seller can add generic superlatives to the description (e.g., ``best popular best greatest sunscreen'') to boost the product's rank, but that might make the description sound artificial or spam-like, and hence, easily discernible from genuine descriptions. Similarly, the seller can add text that is semantically consistent with the original description, but it is unclear how that would influence the ranking of the item.
Second, text generation methods are not designed to generate text that maximizes ranking performance in a recommender system. 
As a result, existing generation methods need to be extended to make them ranking-aware. 
Third, since sellers do not have knowledge about the recommender system, attacks should be conducted in a black-box attack setting---i.e., assuming no access to training data and model parameters. 

We propose the \textbf{A}dversarial \textbf{T}ext \textbf{R}ewriting algorithm (\textbf{\method}) to rewrite target items' descriptions that are ranking-optimized as per recommendations, while maintaining utility to human readers. 
\method presents two distinct rewriting strategies, catering to the availability of attack resources and the choice of language models:\\
(1) The first approach \ft involves a \textit{two-phase fine-tuning process}, which proves advantageous when employed in conjunction with relatively small-sized language models, such as \pointer~\cite{zhang2020pointer} or \opt-350M~\cite{zhang2022opt}. 
In the first phase, \ft fine-tunes the language model on the text of the entire recommendation dataset to contextualize the generation to the domain-specific knowledge present in the dataset.
In the second phase, \ft introduces joint learning objectives to generate text that simultaneously optimizes item ranking and content quality.

(2) The second approach \icl, \textit{in-context learning} (ICL) with carefully crafted prompts, is tailored for widely available LLMs such as \llama~\cite{touvron2023llama}, GPT-4~\cite{openai2023gpt4}. While the two-phase fine-tuning approach generates more effective attacks, in-context learning facilitates exploitation of the impressive text generation capabilities of LLMs. 
The ICL approach rewrites a product description through sophisticated prompt engineering with few adversarial examples obtained from the small language model.

Our experiments on 4 text-aware recommender systems and 3 metadata-rich datasets  show that the recommenders exhibit vulnerability against text rewriting attacks. Specifically, the item descriptions generated by \method boost the ranking of target items significantly compared to its original ranking, outperforming 7 competitive baselines. 
Notably, we find that the in-context learning of LLMs exhibits slightly worse attack performance but generates more fluent text, compared to the fine-tuning approach.
Qualitative evaluations of the rewritten text created by \method confirm better readability and accuracy compared to a baseline. 
We release the anonymized code and dataset for reproducibility.~\footnote{\url{https://github.com/sejoonoh/ATR/}}

In summary, the main contributions of our paper are:

\begin{itemize}
    \item 
    We propose the text description rewriting attack and study the adversarial robustness of text-aware recommender systems against the rewriting attack. 
    \item We propose a two-phase fine-tuning method to rewrite descriptions of target item(s). Our approach, \ft, introduces a rank promotion loss to optimize item ranking and works in both black- and white-box settings. 
    \item We propose a novel in-context learning method \icl for the text rewriting attack based on LLMs such as \llama. Using curated prompts with few examples, we obtain fluent and rank-boosting rewritten product descriptions.
\end{itemize}

%% file: 020related_work.tex
\textbf{Text-Aware and Multimodal Recommenders.}
Conventional recommenders~\cite{hochreiter1997long, kang2018self, li2020time} have suffered from cold-start problems, where users/items with no or few interactions are given poor recommendations. To address those issues, multimodal recommenders~\cite{truong2021multi, di2020taamr, salah2020cornac, tang2019adversarial, sun2020multi} have been proposed, which incorporate additional features such as text and image of an item into the models.  
In \cite{truong2021multi}, the authors describe important modalities such as text~\cite{li2017collaborative, kim2016convolutional}, images~\cite{di2020taamr, he2016vbpr, tang2019adversarial}, and graphs~\cite{sun2020multi,  fan2019graph} that can be used for recommendations. 
Text-aware recommenders~\cite{shalaby2022m2trec, kim2016convolutional, zhou2019content, chuang2020tpr} utilize language models~\cite{kenton2019bert, radford2019language} to vectorize text information (e.g., titles, descriptions). 
Those textual features can be directly fed to the model or combined with existing user and item representations.
They have shown superior prediction performance  compared to conventional methods~\cite{shalaby2022m2trec, chuang2020tpr}, particularly in the cold-start setup.

\noindent \textbf{Targeted Attacks on Recommender Systems.}
Promoting target items has been actively investigated in the domain of recommender systems~\cite{chiang2023shilling, zhang2022pipattack, liu2021adversarial, song2020poisonrec, zhang2020practical, yue2021black, fan2021attacking, di2020taamr, cohen2021black, guo2023targeted, huang2023single, fan2023adversarial}. 
Most of the existing literature~\cite{song2020poisonrec, zhang2020practical, yue2021black, zhang2022pipattack, guo2023targeted, huang2023single} \textit{injects} fake user profiles into the training/test data (i.e., \textit{shilling attacks}). 
Such fake user profiles contain carefully crafted interactions (e.g., clicks) generated by gradient-based algorithms such as FGSM~\cite{Goodfellow2015} to promote the target items. 
Recently, reinforcement learning has been utilized to train the attack agent to perform more effective targeted shilling attacks~\cite{chiang2023shilling, zhang2020practical, song2020poisonrec}.
We do not include shilling attack approaches as baselines since 
most of them cannot generate new adversarial item descriptions, and our goal is to measure the vulnerability of text-aware recommender systems caused by the text modality.

\noindent \textbf{Text Generation/Rewriting Methods.}
Recent approaches for text generation can be categorized as follows: (1) soft-constraint generation, e.g., generating the item description of a laptop, and (2) hard-constraint generation, e.g., generating the description of a laptop using the phrases `512 GB SSD', `Graphics processor', and `16 GB RAM'.
\pointer~\cite{zhang2020pointer}  is a powerful hard constraint-based generation model that leverages BERT~\cite{kenton2019bert}-style language models. 
On the other hand, \opt~\cite{zhang2022opt}, \gptfour~\cite{openai2023gpt4}, \falcon~\cite{falcon40b}, and \llama~\cite{touvron2023llama} are soft-constraint text generators that create new text in an autoregressive manner, i.e., they predict the next word given previous words (the prompt). 
\argkdd~\cite{chiang2023shilling} is a state-of-the-art fake review generation method for target item promotion.  
However, as shown in Table~\ref{tab:comparators}, existing text generation models are not designed to optimize content to perform best as per a recommender system.

%% file: 030preliminaries.tex
\begin{figure*}[t!]
    \centering
    \includegraphics[width=0.7\linewidth]{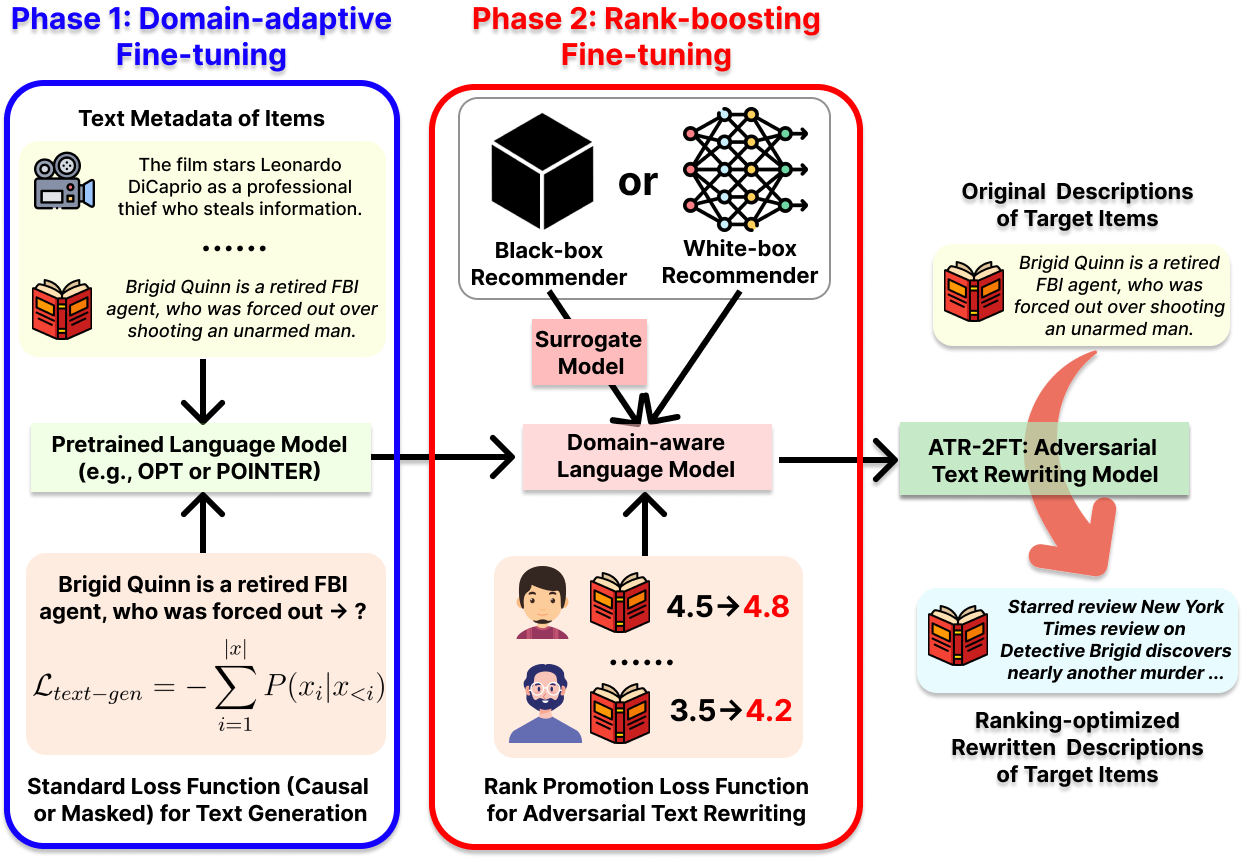}
    \caption{An overview of the two-phase fine-tuning process of \ft for promoting target items in text-aware recommender systems. 
    \ft first fine-tunes a pre-trained text generation model with product descriptions to learn domain-specific knowledge (Phase 1). Next, \ft performs a special fine-tuning of the language model with a rank promotion objective (Phase 2).
    The fully fine-tuned text generation model after Phase 2 can be used to generate ranking-optimized descriptions of target item(s). 
    }
    \label{fig:overview}
\end{figure*}

\noindent \textbf{Text-aware Recommender Systems.}
A text-aware recommender model $\mathcal{M}$ is trained on the interactions between $|\mathcal{U}|$ users and $|\mathcal{I}|$ items. Each item $i$ has an associated text attribute $T_i$, such as its title and description. 
$\mathcal{M}$ learns a prediction/scoring function $\Theta(u,i; \mathcal{M}) \coloneqq \mathcal{M}(E^{\mathcal{U}}_u, E^{\mathcal{I}}_i, E^{\mathcal{T}}_{T_i})$ to score the user-item interaction between $u$ and $i$ based on trainable user embedding  $E^{\mathcal{U}}_u$, trainable item embedding $E^{\mathcal{I}}_i$, and item $i$'s text embedding $E^{\mathcal{T}}_{T_i}$.  
The prediction function $\Theta(u,i; \mathcal{M})$ can be either predicting a rating score (e.g., between 1 and 5)~\cite{zhou2019content, tang2019adversarial} or an interaction probability between the user and item~\cite{hou2022unisrec, zhang2019feature}.
Text embeddings can be computed using language models such as SBERT~\cite{reimers2019sentence}, \opt~\cite{zhang2022opt}, and \llama~\cite{touvron2023llama}.

\noindent \textbf{Attack Setup and Objective.}
Consider the scenario in an e-commerce platform, where a malicious seller (i.e., attacker) wants to promote the ranking of the items she sells, i.e., the target items. 
the attacker can edit the textual descriptions of her target items, but does not have the ability to edit other properties of target items (e.g., price or image) or the descriptions of the other items. 
A recommendation model $\mathcal{M}$ is pre-trained and frozen, so the attacker also cannot manipulate $\mathcal{M}$'s parameters and inject new users, items, or interactions.
The attacker does not know the type of text encoder used by $\mathcal{M}$. 
The attacker can access and download item metadata (e.g., description) from public websites (e.g., Amazon data~\cite{ni2019justifying}) via crawling, which can be used for fine-tuning language models later.

Given a target item set $\mathcal{G}$, the attacker's goal is to rewrite the description $T_i$ to $\hat{T}_i$ of each target item $\forall i \in \mathcal{G}$ such that 
\begin{equation}
    \Theta(u, i, T_i; \mathcal{M}) < \Theta(u, i, \hat{T}_i; \mathcal{M}), \forall u \in \mathcal{U} 
\end{equation}
The attacker also aims to maintain semantic similarity between the original and rewritten text. i.e., $S(T_i,\hat{T}_i)$ should be high where $S$ is a similarity function (e.g., cosine similarity) between two texts. 

\noindent \textbf{Attack Scenarios.}
We consider the following attack scenarios:
\begin{enumerate}
    \item {\textit{Black-box Attack.}} In a black-box setup, the attacker does not have access to the recommendation model's parameters and its training data, while the attacker can obtain a ranked list of items or a rank of a specific item from the recommender.
    \item \textit{White-box Attack.} E-commerce service providers (e.g., Amazon) have full access to their recommendation model and training data. 
    They want to red team against their recommender system and estimate the \textit{worst-case} stability of their recommender system against the text rewriting attack. 
\end{enumerate}

We use three different types of pre-trained text generation models as backbones: POINTER~\cite{zhang2020pointer}, OPT~\cite{zhang2022opt}, and LLaMA-2~\cite{touvron2023llama}.

%% file: 040proposed_method.tex
Our proposed method \method offers two different rewriting approaches depending on the size of the text generation model and whether it is open-source. The first one is \ft, a two-phase fine-tuning of the text generation model that is ideal for relatively small-sized and open-source language models such as \pointer~\cite{zhang2020pointer} and \opt-350M~\cite{zhang2022opt} (see Section~\ref{sec:method:fine_tuning}). 
The second one is \icl, in-context learning (ICL) with curated prompts that is suitable for LLMs such as \llama~\cite{touvron2023llama} and \falcon~\cite{falcon40b} and proprietary models such as \gptfour~\cite{openai2023gpt4} (see Section~\ref{sec:method:ICL}), since it does not require any fine-tuning or training of such LLMs. 
The fine-tuning demonstrates superior attack performance compared to ICL, while ICL generates more fluent and realistic text by leveraging the LLMs. 

\subsection{Proposed Method: \ft}
\label{sec:method:fine_tuning}
When the attacker has enough computational resources for fine-tuning a language model, two-phase fine-tuning is an effective approach to produce ranking-optimized target item descriptions. \ft fine-tunes a text generation method such as \pointer~\cite{zhang2020pointer} and \opt~\cite{zhang2022opt} in two phases. 
After the fine-tuning, \ft can generate rank-boosting descriptions of target items.

\subsubsection{Phase 1: Fine-tuning for Domain Adaptation}
\label{sec:method:phase1}

The first phase fine-tunes the pre-trained text generation model on the entire product descriptions of the recommendation dataset, thus inducing the domain adaptation effect.
Note that the item information (e.g., descriptions) required for the fine-tuning is \textbf{easily accessible to attackers} via crawling websites or using APIs provided by online platforms.
Since item descriptions in a recommendation dataset can be notably different from the broader internet data (e.g., Amazon Electronics item descriptions are very niche and specific compared to the content present in the entire internet), not introducing domain adaptation explicitly will lead to text generations that are generic and irrelevant to the item's domain. 
Thus, this phase ensures that the generator language model is able to generate text that is within the dataset domain. 

The loss function for the Phase-1 fine-tuning depends on the language model; for causal language models, the loss function is to maximize the probability of the next token given previous tokens, while masked language models predict a masked token in a sequence by attending all the tokens on the left and right from it.
We represent the Phase 1 fine-tuning loss as $\mathcal{L}_{text-gen}$.

\subsubsection{Phase 2: Rank-boosting Fine-tuning with Rank Promotion Loss}
\label{sec:method:phase2}
In Phase 2, we fine-tune the text generation model to produce optimal descriptions for target items via multi-objective fine-tuning.
Specifically, we have two objective functions that are optimized jointly --  the first one is the text generation loss ($\mathcal{L}_{text-gen}$), and the second one is a rank promotion loss we define to maximize the predicted scores of target items among all users. 

\noindent \textbf{Rank Promotion Loss for Promoting Target Items.}
The rank promotion loss aims to increase the prediction scores of target items for each user higher than the maximum prediction score given by the user (to any item). The prediction scores are generated by a text-aware recommender $\mathcal{M}$.
In this way, the target items would be ranked at the top of each user's recommendation list. 
Specifically, the proposed rank promotion loss $\mathcal{L}_{promotion}$ is as follows.
\begin{equation}
\begin{split}
\mathcal{L}_{promotion} = \sum_{u \in \mathcal{U'}} \sum_{i \in \mathcal{G'}} \max(\mathcal{R}^{u}_{max}-\Theta(u,i; \mathcal{M})+\lambda_1, 0), \\
\mathcal{R}^{u}_{max} = \max_{\forall i \in \mathcal{I}} \{\Theta(u,i; \mathcal{M})\},
\end{split}
\label{eq:promotion_loss}
\end{equation}
\noindent where $\mathcal{U'}$ and $\mathcal{G'}$ are randomly-sampled subsets of all users $\mathcal{U}$ and all target items $\mathcal{G}$, respectively.
In a black-box setting, we generate fake user profiles $\mathcal{Z}$ and replace $\mathcal{U}$ with $\mathcal{Z}$.
For faster training, we use sampled sets of users and target items instead of the entire sets. We use different subsets of users and target items for each training batch to ensure the promotion is effective across all users. The rank promotion loss \textbf{does not} require original user-item interactions. 
$\mathcal{R}^{u}_{max}$ is defined as the maximum prediction score given by a user $u$ to any item $i \in \mathcal{I}$, as predicted by $\mathcal{M}$. 
$\mathcal{R}^{u}_{max}$ is updated every fine-tuning epoch. It can be computed quickly using the inference mode of $\mathcal{M}$.
Finally, $\lambda_1$ is a margin coefficient (a hyperparameter). 
\textbf{Maximizing $\Theta(u,i; \mathcal{M})$ directly} instead of comparing with $\mathcal{R}^{u}_{max}$ in Eq.~\eqref{eq:promotion_loss} can lead to overfitting of the text generation model since it can unnecessarily try to increase the prediction score of a target item even if it is already top-ranked for a user.

Calculating the rank promotion loss requires access to $\Theta(u,i; \mathcal{M})$ of $\mathcal{M}$ and its gradients. In a white-box setting, these are directly accessible to the attacker, while they are not available in a black-box setting. Thus, the attacker needs to employ \textbf{a model surrogation} to create a local recommender $\mathcal{M'}$ that \textbf{mimics the original one $\mathcal{M}$}. We find the surrogate model $\mathcal{M'}$ using the below loss function~\eqref{eq:surrogate_loss} which minimizes prediction score differences between $\mathcal{M}$ and $\mathcal{M'}$.
\begin{equation}
\mathcal{L}_{surrogate} = \sum_{\forall (u,i) \in \mathcal{X'}_{tr}} \{\Theta(u,i; \mathcal{M})-\Theta(u,i; \mathcal{M'})\}^2
\label{eq:surrogate_loss}
\end{equation}
Thus, in a black-box setting, $\mathcal{M'}$ is used in place of $\mathcal{M}$. We further describe the details of the black-box model surrogation in Appendix.

\noindent \textbf{Rank-boosting Fine-tuning of the Text Generation Model.}
\ft adds the rank promotion loss as a form of  regularization to the text generation loss $\mathcal{L}_{text-gen.}$ and optimizes both losses jointly on product description data, as follows.
\begin{equation}
\mathcal{L}_{\ft} = \mathcal{L}_{text-gen.} + \lambda \mathcal{L}_{promotion}, 
\label{eq:main_loss}
\end{equation}
\noindent where $\lambda$ is a regularization strength (a hyperparameter).
\ft can fine-tune a text generation model to produce ranking-optimized item descriptions (due to $\mathcal{L}_{promotion}$), while maintaining the fluency and semantic meaning (due to $\mathcal{L}_{text-gen.}$).  
We can avoid overfitting and catastrophic forgetting during the fine-tuning using small learning rate and few fine-tuning epochs (e.g., early stopping).

Note that the fine-tuning process does not involve text generation and the loss~\eqref{eq:main_loss} is \textbf{differentiable}. Moreover, the \textbf{main trainable parameters} while computing the loss~\eqref{eq:main_loss} are the text generation model's parameters and text embeddings of items $E^\mathcal{T}$, as \textbf{the recommender $\mathcal{M}$ is pre-trained and frozen}.
We use the same language model being used for the fine-tuning to compute the text embeddings $E^\mathcal{T}$. 
Minimizing the promotion loss~\eqref{eq:promotion_loss} will guide the text generation model to generate ranking-optimized text. 
Concurrently, the text generation loss in Eq.\eqref{eq:main_loss} will balance the parameter updates to create fluent and relevant rewritten text.

\begin{figure}[t!]
    \centering
    \includegraphics[width=0.8\linewidth]{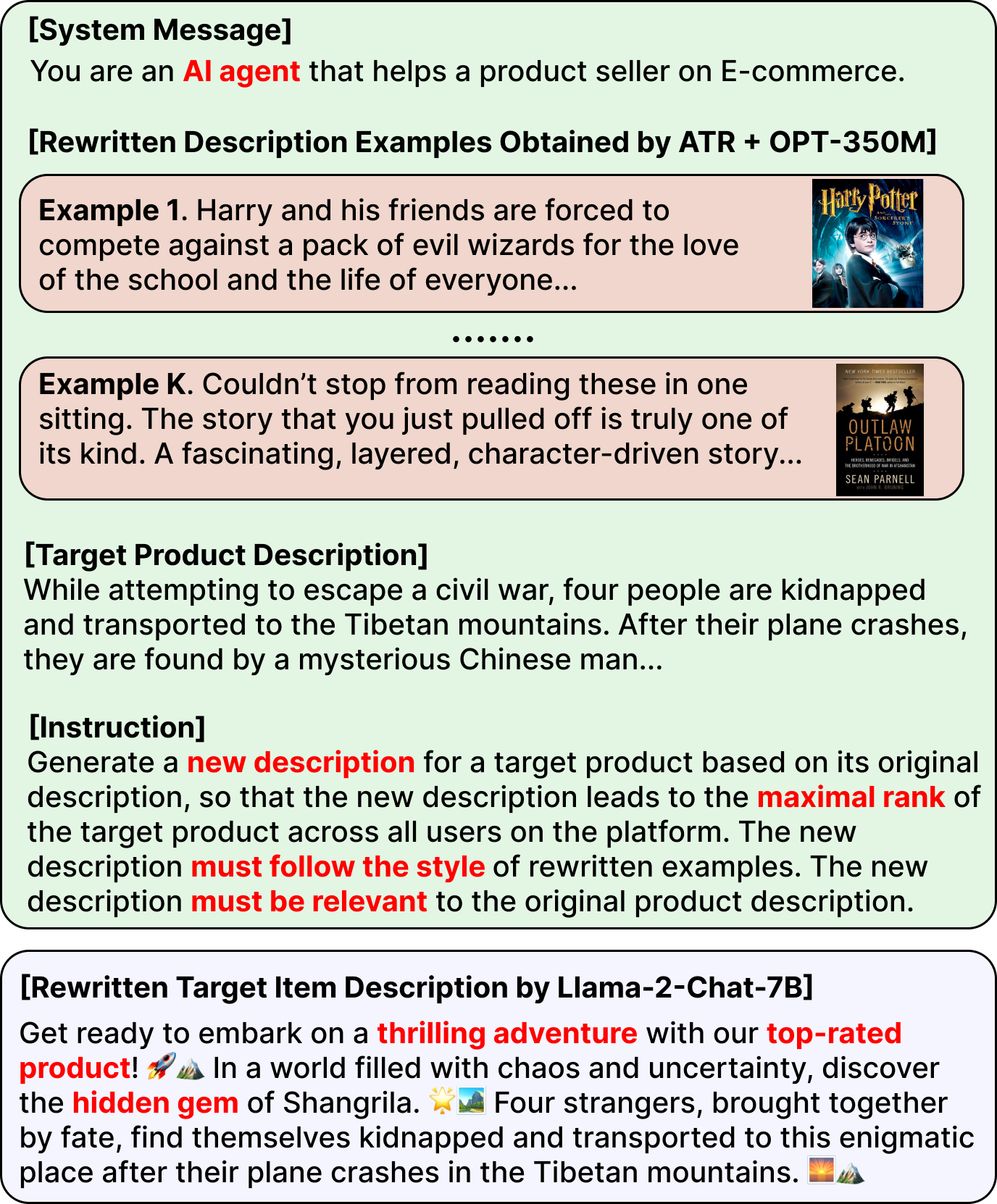}
    \caption{A prompt example for our proposed in-context learning for the \llama-Chat-7B language model. 
    }
    \label{fig:in_context_learning}
\end{figure}

\subsection{Proposed Method: \icl}
\label{sec:method:ICL}
Recent advances in generative artificial intelligence have brought larger  and larger pre-trained language models  such as \llama~\cite{touvron2023llama} and GPT-4~\cite{openai2023gpt4}. 
Applying the proposed two-phase fine-tuning on such LLMs can be computationally prohibitive for attackers, as the fine-tuning updates all parameters of the LLM. 
An alternative approach is to employ in-context learning (ICL)~\cite{dong2023survey} without the need for complete fine-tuning of the LLM. 

The ICL consists of specific instructions for the language model and few examples related to the instructions that the language model can learn from.
For instance, Figure~\ref{fig:in_context_learning} shows an example of a curated prompt used for in-context learning of the \llama-Chat-7B model. 
The prompt starts with a system message stating the general environment and conditions for the language model. It is important to note that we cannot explicitly mention the adversarial attack setup in the prompt as it will be filtered by the LLM easily. 

After the system message, we append few (e.g., 5 or 10) rewritten descriptions generated by the two-phase fine-tuning (\ft) with a smaller language model such as \opt-350M~\cite{zhang2022opt}. These examples act as a guide to the LLM and let the LLM exploit the patterns of adversarially rewritten examples generated by \ft. Note that the ICL is not heavily reliant on \ft, as the fine-tuning of a smaller language model with \ft is needed \textbf{just once} to obtain few rewritten examples, and we can promptly save and load the rewritten examples without extra fine-tuning.
We note that the ICL will be less effective if the examples are not given (i.e., zero-shot learning), or heuristic-based examples (e.g., paraphrasing) are used, rather than the ranking-optimized examples from \ft. 

The final part of the prompt is providing detailed tasks and input data to the language model. We offer the target product description and highlight the rewriting task again. We also put additional constraints on the rewriting: (1) the new text must follow the writing style of the provided examples, and (2) the new text must be relevant to the original one to prevent the hallucination effect. 

For each target item, we randomly sample different rewritten examples, feed the final prompt to the language model, and obtain the rewritten description.
With the carefully-crafted prompt, the language model can quickly learn the essence of rewritten samples and generate new text that is based on the original product description but with more rank-boosting phrases.
Compared to the \ft, \icl produces more fluent and realistic text as it utilizes LLMs (e.g., \llama)  for text generation, while \ft uses relatively small language models (e.g., BERT) for text generation.

%% file: 050experiment.tex
\subsection{Experimental Setup}
\subsubsection{Dataset}
\label{sec:exp:dataset}
We use three recommendation datasets (see Table~\ref{tab:dataset}): \textbf{Amazon Book}, \textbf{Amazon Electronics}~\cite{ni2019justifying},
and \textbf{MovieLens}~\cite{harper2015movielens} which  
contain text descriptions of items. 
We filter out cold-start users and items (e.g., having less than 10 interactions) and items with short descriptions (e.g., less than 50 words).
Among different genres in the Amazon review dataset~\cite{ni2019justifying}, we chose the Book and Electronics genres since they are the largest ones, and their descriptions can be more important for ranking and recommendations than other genres (e.g., the attack may focus on the product image in the Clothing genre). 
We randomly split the interaction data into 81\%/9\%/10\% for training/validation/test sets, respectively.
As the last part of item descriptions often contains noise or less relevant information (e.g., copyright information or seller information), we use the first 100 words of each item description. 
\begin{table}[t!]
\small
	\centering
	\caption{Dataset Statistics. M=million, K=thousand.
	}
\vspace{-3mm}
 	\begin{tabular}{ c | c | c | c | c }
		\toprule
		\textbf{Name} & \textbf{Users} & \textbf{Items} & \textbf{Interactions} & \textbf{\begin{tabular}[c]{@{}c@{}}Avg. Text \\ Length \end{tabular}}\\
		\midrule
		Amazon Book & 28K & 10K &  1.2M & 360 words \\
        Amazon Elec. & 51K & 16K &  589K & 305 words \\
		MovieLens & 78K & 2K & 4.7M & 78 words\\
	\bottomrule
	\end{tabular}	
	\label{tab:dataset}
\end{table}

\subsubsection{Text-aware Recommenders}
\label{sec:exp:recommenders}
We test \method on the state-of-the-art text-aware recommenders. We focus on ``text-as-context'' recommenders that employ text as key input features for predictions; \method is less effective on ``text-as-regularizer'' recommenders~\cite{sachdeva2020useful, mcauley2013hidden} that mainly use text for regularization during training.

\noindent \textbf{\hmf~\cite{zhou2019content} and \amr~\cite{tang2019adversarial}}: The state-of-the-art text-aware collaborative filtering-based recommendation models. 

\noindent    \textbf{\unisrec~\cite{hou2022unisrec}, \fdsa~\cite{zhang2019feature}}: The state-of-the-art text-aware sequential recommendation models.

\subsubsection{Baseline Text Attack/Rewriting Methods}
\label{sec:exp:baselines}
There is no existing text attack/rewriting algorithm that can generate ranking-optimized product descriptions. 
Thus, we employ the following representative textual adversarial attack and generation algorithms that are widely used in the NLP domain. 

\noindent \textbf{\atwot~\cite{yoo2021towards}, \bae~\cite{garg2020bae}, \checklist~\cite{ribeiro2020beyond}}: The state-of-the-art adversarial attack algorithms for NLP classification models. 
As they are designed for binary classifications, we binarize the ratings of our datasets using the average rating as a threshold (e.g., below average $\rightarrow$ 0) and apply the attacks on a BERT-based classifier to find optimal word replacements. 

\noindent \textbf{\gpt~\cite{radford2019language} and \gpt (Fine-tuned)}: a pre-trained language model on English using a causal language modeling. Given a prompt, \gpt can generate realistic and coherent text. We use the original item descriptions as prompts and only use the generated text from \gpt. \gpt (Fine-tuned) indicates \gpt is fine-tuned on our recommendation datasets before generations.

\noindent \textbf{\llama~\cite{touvron2023llama} and GPT-4~\cite{openai2023gpt4}}: zero-shot learning rewriting methods based on \llama-Chat-7B and GPT-4 models. They rewrite a given product description without any rewritten examples.

\subsubsection{Implementation and Hyperparameters}
\label{sec:exp:setup}
\method is implemented in Python and PyTorch and tested in the NVIDIA DGX machine with 5 NVIDIA A100 GPUs.
Default hyperparameters of \method are found by extensive grid searches across datasets. Detailed hyperparameter choices and analyses are given in Section~\ref{sec:exp:hyperparameters}.
The learning rate is set to $10^{-5}$, and the fine-tuning batch size is set to $16$.
The numbers of Phase 1 and Phase 2 epochs are set to 10 and 2, respectively.
We randomly sample 1\% of the total items as target items. The size of sampled users and items $|\mathcal{U'}|$ and $|\mathcal{G'}|$ in Eq.~\eqref{eq:promotion_loss} are set to $0.1|\mathcal{U}|$ and $16$, respectively.
$\lambda$ in  Eq.~\eqref{eq:promotion_loss} is set to $1.0$. 
$\lambda_1$ in  Eq.~\eqref{eq:promotion_loss} is set to $0.01$.
To obtain the text embedding during the test time, we use the state-of-the-art algorithm called SBERT~\cite{reimers2019sentence}.
For \pointer~\cite{zhang2020pointer} and SBERT, we use the all-MiniLM-L6-v2     architecture as a pre-trained language model.
For \opt~\cite{zhang2022opt}, we use the OPT-350m architecture as a pre-trained language model. 

\subsubsection{Evaluation Metrics}
\label{sec:exp:metrics}
The \textbf{primary metric} for \method and baselines is the \textbf{average predicted rank} of target items across all users calculated with and without adversarially rewritten text.  Appear@20 is the secondary metric. To compute Appear@20, we first count how many target items exist in top-20 recommendations for each user; then, we aggregate the counts across all users and divide by $20|\mathcal{U}|$.
We estimate the quality of rewritten text using cosine similarity between the sentence embeddings of original and new item descriptions, Perplexity~\cite{miaschi2021makes} as well as METEOR~\cite{banerjee2005meteor} and BertScore~\cite{zhang2019bertscore} metrics. 
We repeat each experiment 3 times with different random seeds and report average values across 3 runs.
To measure statistical significance, we use the one-tailed $t$-test.

\begin{table*}[t!]
\caption{Effectiveness of our proposed attacks (\ft and \icl) and baselines in promoting the ranks of target items on diverse text-aware recommender systems. \ft promotes the ranks of target items the most via the two-phase fine-tuning process and in-context learning, as per most ranking metrics, while \icl generates more fluent and realistic text than \ft. W.R. = Word Replacements.
} 
\label{tab:amazon_dense}
\small
\vspace{-3mm}
\begin{tabular}{|c|cc|cc|cc|cc}
\hline
\textbf{Recommender Models} &
  \multicolumn{2}{c|}{\begin{tabular}[c]{@{}c@{}}\textbf{\hmf} \cite{zhou2019content}\end{tabular}} &
  \multicolumn{2}{c|}{\begin{tabular}[c]{@{}c@{}}\textbf{\amr} \cite{tang2019adversarial}\end{tabular}} &
  \multicolumn{2}{c|}{\begin{tabular}[c]{@{}c@{}}\textbf{\unisrec} \cite{hou2022unisrec}\end{tabular}} & 
\multicolumn{2}{c|}{\begin{tabular}[c]{@{}c@{}}\textbf{\fdsa} \cite{zhang2019feature}\end{tabular}}
  \\ \hline
  \textbf{\begin{tabular}[c]{@{}c@{}}Evaluation Metrics for \\ Promoting Target Items\end{tabular}} &
  \multicolumn{1}{c|}{\begin{tabular}[c]{@{}c@{}}Pred. \\ Rank\end{tabular} $\boldsymbol{\Big\downarrow}$} &
  \multicolumn{1}{c|}{\begin{tabular}[c]{@{}c@{}}Appear \\ @20\end{tabular} $\boldsymbol{\Big\uparrow}$} &
  \multicolumn{1}{c|}{\begin{tabular}[c]{@{}c@{}}Pred. \\ Rank\end{tabular} $\boldsymbol{\Big\downarrow}$} &
  \multicolumn{1}{c|}{\begin{tabular}[c]{@{}c@{}}Appear \\ @20\end{tabular} $\boldsymbol{\Big\uparrow}$} &
  \multicolumn{1}{c|}{\begin{tabular}[c]{@{}c@{}}Pred. \\ Rank\end{tabular} $\boldsymbol{\Big\downarrow}$} &
  \multicolumn{1}{c|}{\begin{tabular}[c]{@{}c@{}}Appear \\ @20\end{tabular} $\boldsymbol{\Big\uparrow}$} &
   \multicolumn{1}{c|}{\begin{tabular}[c]{@{}c@{}}Pred. \\ Rank\end{tabular} $\boldsymbol{\Big\downarrow}$} &
  \multicolumn{1}{c|}{\begin{tabular}[c]{@{}c@{}}Appear \\ @20\end{tabular} $\boldsymbol{\Big\uparrow}$} 
  \\ \hline
\textbf{Original Descriptions} &
  \multicolumn{1}{c|}{\cellcolor{red!25} 542.5} &
  \multicolumn{1}{c|}{\cellcolor{red!25} 0.1093} &
  \multicolumn{1}{c|}{\cellcolor{red!25} 549.0} &
  \multicolumn{1}{c|}{\cellcolor{red!25} \textbf{0.0909}} &
  \multicolumn{1}{c|}{\cellcolor{red!25} 552.1} &
  \multicolumn{1}{c|}{\cellcolor{red!25} 0.0880} &
  \multicolumn{1}{c|}{\cellcolor{red!25} 550.4} &
  \multicolumn{1}{c|}{\cellcolor{red!25} 0.0971}
   \\ \hline
\textbf{\begin{tabular}[c]{@{}c@{}}\atwot W.R. \end{tabular}} &
  \multicolumn{1}{c|}{512.1} &
  \multicolumn{1}{c|}{0.1975} &
  \multicolumn{1}{c|}{551.8} &
  \multicolumn{1}{c|}{0.0850} &
  \multicolumn{1}{c|}{550.6} &
  \multicolumn{1}{c|}{0.0924} &
  \multicolumn{1}{c|}{547.9} &
  \multicolumn{1}{c|}{0.1018} 
   \\ 
\textbf{\begin{tabular}[c]{@{}c@{}}\bae W.R. \end{tabular}} &
  \multicolumn{1}{c|}{526.9} &
  \multicolumn{1}{c|}{0.1624} &
  \multicolumn{1}{c|}{551.6} &
  \multicolumn{1}{c|}{0.0884} &
  \multicolumn{1}{c|}{546.5} &
  \multicolumn{1}{c|}{0.0971} &
  \multicolumn{1}{c|}{547.0} &
  \multicolumn{1}{c|}{0.1034}
   \\
\textbf{\begin{tabular}[c]{@{}c@{}}\checklist W.R. \end{tabular}} &
  \multicolumn{1}{c|}{546.7} &
  \multicolumn{1}{c|}{0.1266} &
  \multicolumn{1}{c|}{550.7} &
  \multicolumn{1}{c|}{0.0886} &
  \multicolumn{1}{c|}{554.1} &
  \multicolumn{1}{c|}{0.0886} &
  \multicolumn{1}{c|}{548.7} &
  \multicolumn{1}{c|}{0.0998}
   \\ 
\textbf{\begin{tabular}[c]{@{}c@{}}\gpt Generation \end{tabular}} &
  \multicolumn{1}{c|}{650.8} &
  \multicolumn{1}{c|}{0.1625} &
  \multicolumn{1}{c|}{544.8} &
  \multicolumn{1}{c|}{0.0663} &
  \multicolumn{1}{c|}{582.1} &
  \multicolumn{1}{c|}{0.0491} &
  \multicolumn{1}{c|}{543.9} &
  \multicolumn{1}{c|}{0.0991}
   \\
\textbf{\begin{tabular}[c]{@{}c@{}}\gpt (Fine-tuned) \end{tabular}} &
  \multicolumn{1}{c|}{533.0} &
  \multicolumn{1}{c|}{0.2795} &
  \multicolumn{1}{c|}{549.9} &
  \multicolumn{1}{c|}{0.0722} &
  \multicolumn{1}{c|}{597.6} &
  \multicolumn{1}{c|}{0.0317} &
  \multicolumn{1}{c|}{546.3} &
  \multicolumn{1}{c|}{0.1003}
   \\   
\textbf{\begin{tabular}[c]{@{}c@{}}\llama Zero-shot \end{tabular}} &
  \multicolumn{1}{c|}{635.3} &
  \multicolumn{1}{c|}{0.0709} &
  \multicolumn{1}{c|}{549.3} &
  \multicolumn{1}{c|}{0.0895} &
  \multicolumn{1}{c|}{546.2} &
  \multicolumn{1}{c|}{0.0930} &
  \multicolumn{1}{c|}{544.7} &
  \multicolumn{1}{c|}{0.0921}
   \\   
\textbf{\begin{tabular}[c]{@{}c@{}}GPT-4 Zero-shot \end{tabular}} &
  \multicolumn{1}{c|}{579.4} &
  \multicolumn{1}{c|}{0.1230} &
  \multicolumn{1}{c|}{542.6} &
  \multicolumn{1}{c|}{0.0848} &
  \multicolumn{1}{c|}{522.8} &
  \multicolumn{1}{c|}{0.0896} &
  \multicolumn{1}{c|}{549.7} &
  \multicolumn{1}{c|}{0.1038}
   \\   \hline
   \textbf{\begin{tabular}[c]{@{}c@{}}\ft (\opt; white-box)\end{tabular}} &
  \multicolumn{1}{c|}{\cellcolor{blue!25} \textbf{366.0}} &
  \multicolumn{1}{c|}{\cellcolor{blue!25} \textbf{0.5591}} &
  \multicolumn{1}{c|}{\cellcolor{blue!25} \textbf{497.1}} &
  \multicolumn{1}{c|}{\cellcolor{blue!25} 0.0897} &
  \multicolumn{1}{c|}{\cellcolor{blue!25} \textbf{394.8}} &
  \multicolumn{1}{c|}{\cellcolor{blue!25} \textbf{0.1801}} &
  \multicolumn{1}{c|}{\cellcolor{blue!25} \textbf{514.7}} &
  \multicolumn{1}{c|}{\cellcolor{blue!25} \textbf{0.1296}}
   \\ 
\textbf{\begin{tabular}[c]{@{}c@{}}\ft (\pointer; white-box)\end{tabular}} &
  \multicolumn{1}{c|}{\cellcolor{blue!25} 464.7} &
  \multicolumn{1}{c|}{\cellcolor{blue!25} 0.4481} &
  \multicolumn{1}{c|}{\cellcolor{blue!25} 538.3} &
  \multicolumn{1}{c|}{\cellcolor{blue!25} 0.0722} &
  \multicolumn{1}{c|}{\cellcolor{blue!25} 443.3} &
  \multicolumn{1}{c|}{\cellcolor{blue!25} 0.1069} &
  \multicolumn{1}{c|}{\cellcolor{blue!25} 541.4} &
  \multicolumn{1}{c|}{\cellcolor{blue!25} 0.1060}
   \\ 
    \textbf{\begin{tabular}[c]{@{}c@{}}\ft (\opt; black-box)\end{tabular}} &
  \multicolumn{1}{c|}{\cellcolor{blue!25} 385.0} &
  \multicolumn{1}{c|}{\cellcolor{blue!25} 0.5519} &
  \multicolumn{1}{c|}{\cellcolor{blue!25} 529.2} &
  \multicolumn{1}{c|}{\cellcolor{blue!25} 0.0821} &
  \multicolumn{1}{c|}{\cellcolor{blue!25} 456.5} &
  \multicolumn{1}{c|}{\cellcolor{blue!25} 0.1471} &
  \multicolumn{1}{c|}{\cellcolor{blue!25} 540.1} &
  \multicolumn{1}{c|}{\cellcolor{blue!25} 0.1044}
   \\ 
   \textbf{\begin{tabular}[c]{@{}c@{}}\ft (\pointer; black-box)\end{tabular}} &
  \multicolumn{1}{c|}{\cellcolor{blue!25} 380.7} &
  \multicolumn{1}{c|}{\cellcolor{blue!25} 0.5528} &
  \multicolumn{1}{c|}{\cellcolor{blue!25} 527.3} &
  \multicolumn{1}{c|}{\cellcolor{blue!25} 0.0697} &
  \multicolumn{1}{c|}{\cellcolor{blue!25} 523.7} &
  \multicolumn{1}{c|}{\cellcolor{blue!25} 0.1135} &
  \multicolumn{1}{c|}{\cellcolor{blue!25} 550.1} &
  \multicolumn{1}{c|}{\cellcolor{blue!25} 0.0907}
   \\
   \textbf{\begin{tabular}[c]{@{}c@{}}\icl \end{tabular}} &
  \multicolumn{1}{c|}{\cellcolor{blue!25} 536.1} &
  \multicolumn{1}{c|}{\cellcolor{blue!25} 0.3573} &
  \multicolumn{1}{c|}{\cellcolor{blue!25} 526.1} &
  \multicolumn{1}{c|}{\cellcolor{blue!25} 0.0798} &
  \multicolumn{1}{c|}{\cellcolor{blue!25} 433.0} &
  \multicolumn{1}{c|}{\cellcolor{blue!25} 0.1624} &
  \multicolumn{1}{c|}{\cellcolor{blue!25} 534.9} &
  \multicolumn{1}{c|}{\cellcolor{blue!25} 0.1017}
   \\ \hline
   \textbf{Rel. \% Improvements} &
  \multicolumn{1}{c|}{\cellcolor{blue!25} \textbf{+28.5\%}} &
  \multicolumn{1}{c|}{\cellcolor{blue!25} \textbf{+100.0\%}} &
  \multicolumn{1}{c|}{\cellcolor{blue!25} \textbf{+9.2\%}} &
  \multicolumn{1}{c|}{\cellcolor{blue!25} \textbf{-1.3\%}} &
  \multicolumn{1}{c|}{\cellcolor{blue!25} \textbf{+24.5\%}} &
  \multicolumn{1}{c|}{\cellcolor{blue!25} \textbf{+85.5\%}} &
  \multicolumn{1}{c|}{\cellcolor{blue!25} \textbf{+5.4\%}} &
  \multicolumn{1}{c|}{\cellcolor{blue!25} \textbf{+24.9\%}}
   \\ \hline
\end{tabular}
\label{tab:promotion_all}
\end{table*}

\subsection{Effectiveness of \method in Promoting Targets}
\label{sec:exp:effectiveness_fine_tuning}
We first evaluate how much \method and baselines can promote the ranks of target items via rewriting item descriptions across diverse text-aware recommender systems on the Amazon Book dataset. 
We focus on the Amazon Book dataset since the book descriptions contain rich and diverse information, including various styles, and descriptions in the Book genre play a crucial role for users and recommender systems to recommend books.
We also evaluate \method on the Amazon Electronics and MovieLens datasets.

The \textbf{impact of \method on overall recommendation accuracy will be insignificant} as our attack does not change the recommendation model parameters, and descriptions of non-target items will not be affected by our attacks. 
Empirically, the test accuracy (RMSE) of a recommendation model \hmf~\cite{zhou2019content} on Amazon Book dataset with and without \ft is 0.76816 and 0.76821, which verifies that the recommendation quality remains almost the same after the attack. This is true for all models and all datasets.

\subsubsection{Attack Performance Summary}
As shown in Table~\ref{tab:amazon_dense}, \method successfully promotes the ranks of target items in most cases with statistical significance (p-values $<$ 0.01; marked with *), compared to original descriptions and baselines. For instance, across all recommendation models, the white-box attack of \ft with \opt shows $5.4\%-28.5\%$ improvements in the average predicted rank metric (lower is better). We note that relative \% improvements indicate how much \method improves the ranking metrics compared to the best baseline (including the  results of original descriptions).

\icl generates rank-boosting descriptions of target items using in-context learning, but its attack performance is limited versus \ft. It is expected since \ft fully fine-tunes parameters of a language model, while \icl does not. \icl still has merits in text quality (e.g., fluency and coherency; see Section~\ref{sec:exp:qualitative_analysis}). 

Between \opt~\cite{zhang2022opt} and \pointer~\cite{zhang2020pointer} used in \ft, \opt mostly outperforms \pointer in terms of ranking metrics. This can be attributed to the differences between their generation schemes. \pointer is a hard-constrained generation model that reconstructs text using keywords extracted from the original text, while \opt is a soft one that takes an original text as a prompt and generates new text in an autoregressive manner. Thus, manipulating \opt with our rank promotion loss can be easier and more effective than \pointer, as \pointer is forced to preserve original keywords (that might not be ranking-crucial) even after the fine-tuning.

Word replacement baselines such as \atwot, \checklist, and \bae show marginal or no ranking improvements after rewriting, as they cannot rewrite entire item descriptions.
The LLM-based rewriting baselines (e.g., \llama and GPT-4) show limitations due to the lack of a rank promotion loss during its training and fine-tuning.

In summary, the superior performance of \method indicates that the current item descriptions are not ranking-optimal, and text-aware recommenders are vulnerable to adversarial text rewriting attacks.

\subsubsection{Robustness of Recommenders}
Among all recommenders, \amr~\cite{tang2019adversarial} and \fdsa~\cite{zhang2019feature} demonstrate relatively high robustness to being influenced by text rewriting attacks.
In terms of the average predicted rank metric, the \fdsa model shows the least change against our \ft attack.
Moreover, the Appear@20 metric is not increased by \ft on the \amr model.
Intuitively, their high robustness could be attributed to either their training or prediction mechanism. 
For instance, \amr conducts adversarial training against gradient-based perturbations on text features which exposes the model to perturbed inputs and covers some of the possible variations explored by rewriting attacks. In the case of \fdsa, its prediction function is heavily dependent on item-ID embeddings and less dependent on text features of items. Thus, text perturbation attacks cannot impact the final ranking significantly.
\begin{table*}[t!]
\vspace{3mm}
\caption{Effectiveness of \ft, \icl, and baselines for promoting target items on the \hmf model, given Amazon Electronics and MovieLens datasets. \ft is the best attack on both datasets as per all ranking metrics with statistical significance.
}
\label{tab:other_dataset}
\vspace{-3mm}
\begin{subtable}{0.48\textwidth}
\centering
\small
\caption{Amazon Electronics Dataset}
\label{tab:amazon_electronics}
\begin{tabular}{|c|cc}
\hline
\textbf{Recommender Models} &
  \multicolumn{2}{c|}{\textbf{\hmf~\cite{zhou2019content}}}  \\ \hline
  \textbf{\begin{tabular}[c]{@{}c@{}}Evaluation Metrics for \\ Promoting Target Items\end{tabular}} &
  \multicolumn{1}{c|}{\begin{tabular}[c]{@{}c@{}}Pred. \\ Rank\end{tabular} $\boldsymbol{\Big\downarrow}$} &
  \multicolumn{1}{c|}{\begin{tabular}[c]{@{}c@{}}Appear \\ @20\end{tabular} $\boldsymbol{\Big\uparrow}$}
  \\ \hline
\textbf{Original Descriptions} &
  \multicolumn{1}{c|}{\cellcolor{red!25} 871.6} &
  \multicolumn{1}{c|}{\cellcolor{red!25} 0.0745}
   \\ \hline
\textbf{\begin{tabular}[c]{@{}c@{}}\atwot W.R. \end{tabular}} &
  \multicolumn{1}{c|}{755.7} &
  \multicolumn{1}{c|}{0.2408}
   \\ 
\textbf{\begin{tabular}[c]{@{}c@{}}\bae W.R. \end{tabular}} &
  \multicolumn{1}{c|}{816.2}  &
  \multicolumn{1}{c|}{0.1605}
   \\
\textbf{\begin{tabular}[c]{@{}c@{}}\checklist W.R. \end{tabular}} &
  \multicolumn{1}{c|}{816.8} &
  \multicolumn{1}{c|}{0.1396}
   \\ 
\textbf{\begin{tabular}[c]{@{}c@{}}\gpt  Generation \end{tabular}} &
  \multicolumn{1}{c|}{1033.4} &
  \multicolumn{1}{c|}{0.1547}
   \\
\textbf{\begin{tabular}[c]{@{}c@{}}\gpt (Fine-tuned) \end{tabular}} &
  \multicolumn{1}{c|}{1287.8} &
  \multicolumn{1}{c|}{0.0463}
   \\ 
\textbf{\begin{tabular}[c]{@{}c@{}}\llama Zero-shot \end{tabular}} &
  \multicolumn{1}{c|}{782.9} &
  \multicolumn{1}{c|}{0.2463}
   \\
\textbf{\begin{tabular}[c]{@{}c@{}}GPT-4 Zero-shot \end{tabular}} &
  \multicolumn{1}{c|}{760.8} &
  \multicolumn{1}{c|}{0.2617}
   \\
   \hline
\textbf{\ft (\pointer; white-box)} &
  \multicolumn{1}{c|}{\cellcolor{blue!25} 951.7}  &
  \multicolumn{1}{c|}{\cellcolor{blue!25} 0.1607}
   \\ \hline
\textbf{\ft (\opt; white-box)} &
  \multicolumn{1}{c|}{\cellcolor{blue!25} \textbf{539.9}} &
  \multicolumn{1}{c|}{\cellcolor{blue!25} 0.6545}
   \\ \hline
   \textbf{\ft (\pointer; black-box)} &
  \multicolumn{1}{c|}{\cellcolor{blue!25} 945.9} &
  \multicolumn{1}{c|}{\cellcolor{blue!25} 0.1367}
   \\ \hline
   \textbf{\ft (\opt; black-box)} &
  \multicolumn{1}{c|}{\cellcolor{blue!25} 726.5} &
  \multicolumn{1}{c|}{\cellcolor{blue!25} 0.3655} 
   \\ \hline
   \textbf{\icl} &
  \multicolumn{1}{c|}{\cellcolor{blue!25} 614.5} &
  \multicolumn{1}{c|}{\cellcolor{blue!25} \textbf{0.6722}} 
   \\ \hline
    \textbf{Rel. \% Improvements} &
  \multicolumn{1}{c|}{\cellcolor{blue!25} \textbf{+28.6\%}} &
  \multicolumn{1}{c|}{\cellcolor{blue!25} \textbf{+156.9\%}}
   \\ \hline
\end{tabular}
\end{subtable}
\hfill
\begin{subtable}{0.48\textwidth}
\centering
\small
\caption{MovieLens Dataset}
\label{tab:movielens_dense}
\begin{tabular}{|c|ccc}
\hline
\textbf{Recommender Models} &
  \multicolumn{2}{c|}{\textbf{\hmf~\cite{zhou2019content}}}  \\ \hline
  \textbf{\begin{tabular}[c]{@{}c@{}}Evaluation Metrics for \\ Promoting Target Items\end{tabular}} &
  \multicolumn{1}{c|}{\begin{tabular}[c]{@{}c@{}}Pred. \\ Rank\end{tabular} $\boldsymbol{\Big\downarrow}$} &
  \multicolumn{1}{c|}{\begin{tabular}[c]{@{}c@{}}Appear \\ @20\end{tabular} $\boldsymbol{\Big\uparrow}$} 
  \\ \hline
\textbf{Original Descriptions} &
  \multicolumn{1}{c|}{\cellcolor{red!25} 118.4} &
  \multicolumn{1}{c|}{\cellcolor{red!25} 0.1085}
   \\ \hline
\textbf{\begin{tabular}[c]{@{}c@{}}\atwot W.R. \end{tabular}} &
  \multicolumn{1}{c|}{120.7} &
  \multicolumn{1}{c|}{0.1064}
   \\ 
\textbf{\begin{tabular}[c]{@{}c@{}}\bae W.R. \end{tabular}} &
  \multicolumn{1}{c|}{121.7} &
  \multicolumn{1}{c|}{0.1011}
   \\
\textbf{\begin{tabular}[c]{@{}c@{}}\checklist W.R. \end{tabular}} &
  \multicolumn{1}{c|}{119.3} &
  \multicolumn{1}{c|}{0.1049}
   \\ 
\textbf{\begin{tabular}[c]{@{}c@{}}\gpt  Generation \end{tabular}} &
  \multicolumn{1}{c|}{135.0} &
  \multicolumn{1}{c|}{0.0603}
   \\
\textbf{\begin{tabular}[c]{@{}c@{}}\gpt (Fine-tuned) \end{tabular}} &
  \multicolumn{1}{c|}{117.7} &
  \multicolumn{1}{c|}{0.1365}
   \\
   \textbf{\begin{tabular}[c]{@{}c@{}}\llama Zero-shot \end{tabular}} &
  \multicolumn{1}{c|}{123.1} &
  \multicolumn{1}{c|}{0.0893}
   \\
\textbf{\begin{tabular}[c]{@{}c@{}}GPT-4 Zero-shot \end{tabular}} &
  \multicolumn{1}{c|}{107.3} &
  \multicolumn{1}{c|}{0.1276}
   \\
\hline
\textbf{\ft (\pointer; white-box)} &
  \multicolumn{1}{c|}{\cellcolor{blue!25} 102.2}  &
  \multicolumn{1}{c|}{\cellcolor{blue!25} 0.1739}
   \\ \hline
\textbf{\ft (\opt; white-box)} &
  \multicolumn{1}{c|}{\cellcolor{blue!25} \textbf{73.9}}  &
  \multicolumn{1}{c|}{\cellcolor{blue!25} \textbf{0.2795}}
   \\ \hline
   \textbf{\ft (\pointer; black-box)} &
  \multicolumn{1}{c|}{\cellcolor{blue!25} 88.4} &
  \multicolumn{1}{c|}{\cellcolor{blue!25} 0.2049}
   \\ \hline
   \textbf{\ft (\opt; black-box)} &
  \multicolumn{1}{c|}{\cellcolor{blue!25} 92.4} &
  \multicolumn{1}{c|}{\cellcolor{blue!25} 0.1881}
   \\ \hline
   \textbf{\icl} &
  \multicolumn{1}{c|}{\cellcolor{blue!25} 104.8} &
  \multicolumn{1}{c|}{\cellcolor{blue!25} 0.1121} 
   \\ \hline
    \textbf{Rel. \% Improvements} &
  \multicolumn{1}{c|}{\cellcolor{blue!25} \textbf{+31.1\%}} &
  \multicolumn{1}{c|}{\cellcolor{blue!25} \textbf{+104.8\%}}
   \\ \hline
\end{tabular}
\end{subtable}
\end{table*}

\subsubsection{Black-box Attack Performance of \method}
\label{sec:exp:black-box}
We evaluate the black-box attack performance of \method on the Amazon Book dataset. 
We note that \icl is naturally a black-box attack since it does not fine-tune the language model and optimizes the input prompt only. 

For \ft in a black-box attack setting, we measure the accuracy changes between the black-box and the surrogate model to validate the accuracy of the surrogate model. Ideally, the surrogate model should have similar accuracy to the black-box one. 
For sequential recommenders, we measure changes in Appear@50 after surrogation; for conventional recommenders, we track changes in root-mean-square error (RMSE). 
The Appear@50 changes of \unisrec and \fdsa are 13.4\% and 15.0\%
The RMSE changes of \hmf and \amr are 3.8\% and 4.0\%, respectively. 
Considering the small changes (especially RMSE), we assert that the surrogate models mimic the black-box recommenders well. 

In Table~\ref{tab:promotion_all}, the results corroborate that \method can promote target items' ranks even in the black-box attack setup, while consistently showing higher ranking metrics than the results of baselines and original descriptions.
\ft (black-box) exhibits higher ranking metrics than those of \icl due to its two-phase fine-tuning.
The high performance of the black-box attack can be explained by the fact that the text features play an important role (i.e., higher weight) in the surrogate model (i.e., NCF~\cite{he2017neural}) used for the black-box attack.

\subsubsection{\method on Amazon Electronics and MovieLens datasets}
\label{sec:exp:movielens}
We evaluate the attack performance of \method and baselines on the Amazon Electronics and MovieLens datasets with the \hmf~\cite{zhou2019content} recommender. We highlight attack results on the \hmf model since it is designed for rating datasets, while we observe similar trends on other text-aware recommenders. 

Table~\ref{tab:other_dataset} shows the ranking metrics of \method and baselines, given target items.
Similar to the results on the Amazon Book dataset, \method promotes the ranks of target items the most among all methods with statistical significance (all p-values $<$ 0.01) across two datasets. 
As expected, \ft shows similar or higher attack performance compared to \icl. 
These results show that our attack \method can be applied to diverse recommendation domains. 

On the Amazon Electronics dataset, the fine-tuned \gpt model shows inferior attack performance against original descriptions. The potential reason for this result can be the characteristic of the Electronics genre. This domain contains numerous technical terms and specification numbers that are not observed frequently in \gpt's pre-training dataset. Therefore, even with the fine-tuning, \gpt might produce out-of-domain item descriptions. 

\begin{figure}[t!]
    \centering
    \includegraphics[width=0.85\linewidth]{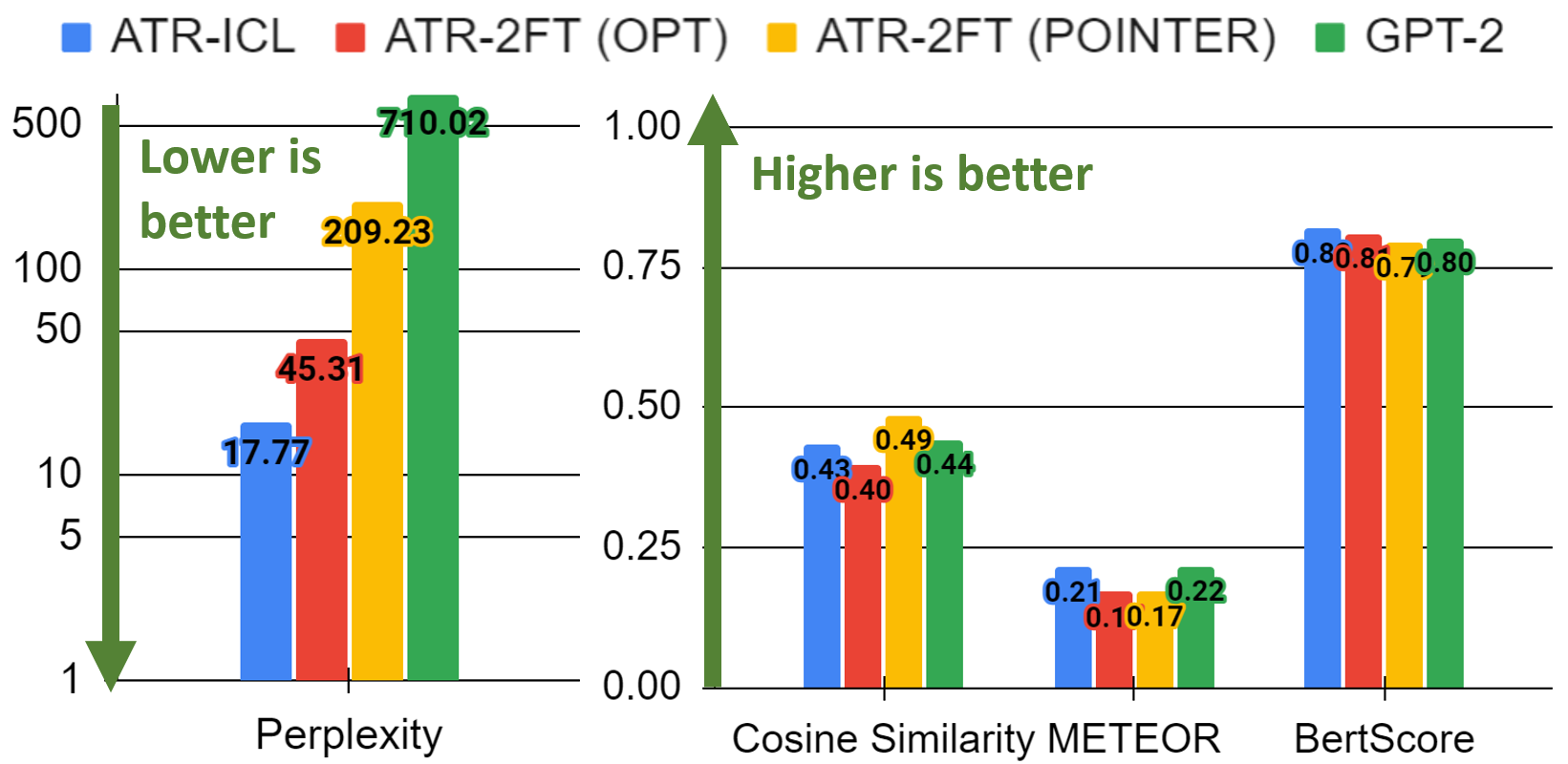}
    \caption{Various text quality metrics of rewritten item descriptions created by \method and the \gpt baseline. }
    \label{fig:text_quality}
\end{figure}

\begin{figure*}[t!]
    \centering
    \includegraphics[width=0.95\linewidth]{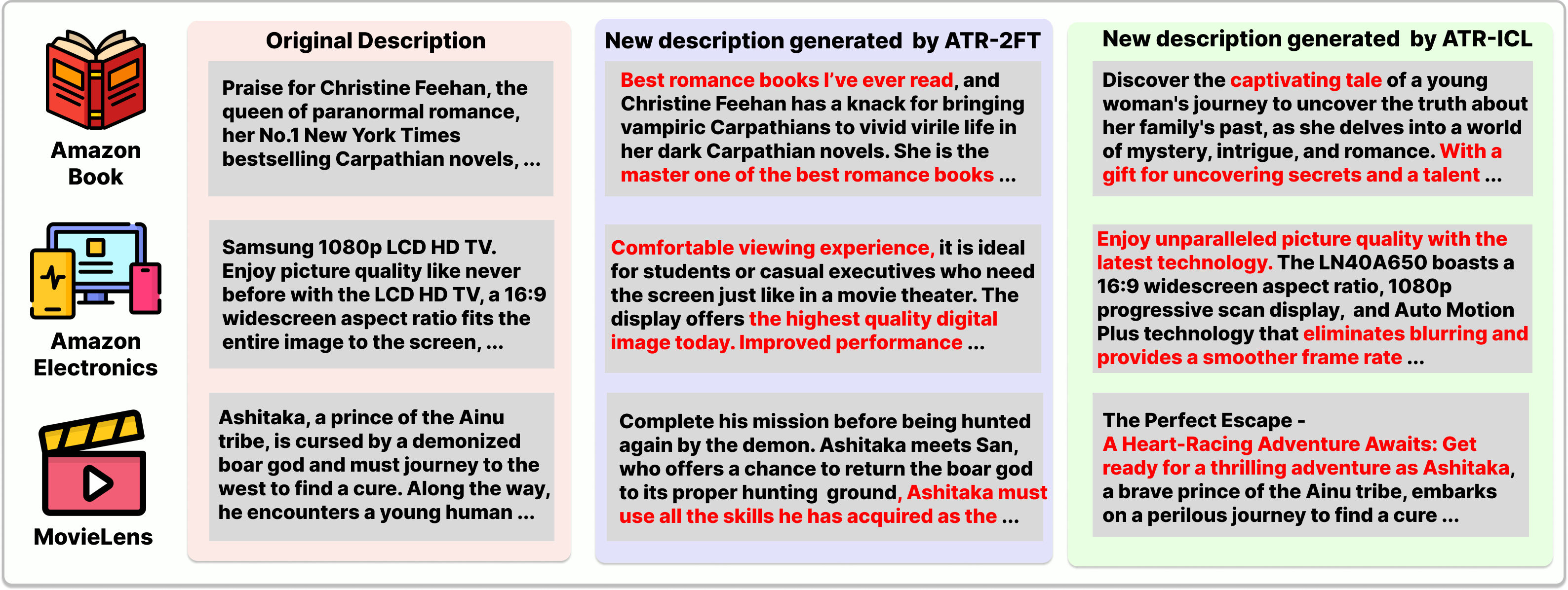}
    \caption{
    Distinctive examples of rewritten item descriptions generated by \ft and \icl on three real-world datasets. The rewritten descriptions are fluent, relevant, and lead to higher ranks of target items than the ranks computed with the original text. 
    }
    \label{fig:case_study}
\end{figure*}

\subsection{Evaluating Rewritten Text Quality}
\label{sec:exp:qualitative_analysis}
We analyze whether the rewritten item descriptions generated by \method are semantically similar and fluent compared to the original descriptions. We do this both using quantitative and qualitative methods. For quantitative test, we use multiple text quality metrics.
For qualitative test, we conduct a human evaluation to compare the text rewriting quality of \method and \gpt models, while also comparing them with the original item descriptions.

The four automated metrics regarding the text quality (Figure~\ref{fig:text_quality}) indicate that \method can generate new text with comparable or better qualities to the text generation baseline \gpt~\cite{radford2019language}.
The high BertScore~\cite{zhang2019bertscore} metric (around 0.8) of \method suggests that \method learns and preserves the semantic meaning of the original item descriptions successfully.
\icl produces more fluent and realistic (i.e., predictable) text than \ft, given its low perplexity score. 
Between \opt and \pointer text generation models, \pointer presents slightly higher coherency but lower fluency, as \pointer can preserve keywords extracted from the original descriptions  while generation and has a smaller model size than \opt.

Figure~\ref{fig:case_study} shows illustrative text rewriting examples of \method across three datasets used in the paper.  
Compared to the original item descriptions (left),
\ft with \opt text generation model (middle) inserts ranking-friendly phrases (colored red) in its rewritten text such as \textit{\textbf{best romance books}} and \textit{\textbf{the highest quality}}, and the rewritten descriptions are still coherent with the original ones.
\icl on \llama-Chat-7B (right) generates the \textbf{most fluent} description with rank-boosting phrases (colored red) across all rewriting methods, as it employs the largest language model. However, it often delivers totally unrelated descriptions as the language model would not be fine-tuned with the given dataset. For instance, on the Amazon Book dataset, \icl's rewriting is about a young woman's journey, which assumes the author of the book and the main character are the same. 
In all cases, ranking metrics of target items are boosted if new adversarial item descriptions are employed. 

Similar to existing literature~\cite{he2021petgen}, we recruit 5 \textbf{human evaluators} to verify the realistic nature of the item descriptions rewritten by \ft. 
Each evaluator was assigned 20 unique tuples of item descriptions, where each tuple contained the original description, the description rewritten by \gpt, and the description rewritten by \ft using \opt. The annotators were instructed to label which item description was more realistic. 
For each description, the corresponding text generation model name is anonymized to prevent potential biases of the evaluator.
We verify that rewritten text generated by \ft is preferred by human evaluators;  the evaluators consider \ft to be better than \gpt and even the Original description 32 times, while \gpt is chosen only 9 times. 

\textbf{Does the hallucination effect matter in an adversarial attack?}
Hallucination~\cite{ji2023survey} in NLP indicates the generated text is plausible but incorrect and nonsensical.
We argue that in an adversarial attack, hallucination is \textit{less relevant} as the primary focus of attackers is to promote target items.
Our qualitative analyses indicate most rewritten descriptions obtained from \method are coherent and accurate; however, some of them contain hallucinations.
One potential way to mitigate hallucination in our attacks is using a hard-constrained text generation approach such as \pointer.

\begin{table}[t!]
\small
\caption{Ablation study of \ft with the \opt generation model on the \hmf recommender and Amazon Book dataset.
}
\begin{tabular}{|c|ccc|}
\hline
\textbf{\begin{tabular}[c]{@{}c@{}}Text Rewriting Methods \\ / Evaluation Metrics \end{tabular} } &
  \multicolumn{1}{c|}{\begin{tabular}[c]{@{}c@{}}Pred. \\ Rank\end{tabular} $\boldsymbol{\Big\downarrow}$} &
   \multicolumn{1}{c|}{\begin{tabular}[c]{@{}c@{}}Pred. \\ Score \end{tabular} $\boldsymbol{\Big\uparrow}$} &
  \multicolumn{1}{c|}{\begin{tabular}[c]{@{}c@{}}Appear \\ @20\end{tabular} $\boldsymbol{\Big\uparrow}$}
    \\ \hline
\textbf{\begin{tabular}[c]{@{}c@{}}Original Descriptions \end{tabular}} &
  \multicolumn{1}{c|}{\cellcolor{red!25} 542.5} &
  \multicolumn{1}{c|}{\cellcolor{red!25} 4.264} &
  \multicolumn{1}{c|}{\cellcolor{red!25} 0.1093}
  \\ \hline
  \textbf{\begin{tabular}[c]{@{}c@{}}\ft only with Phase-1 \end{tabular}} &
   \multicolumn{1}{c|}{550.5} &
   \multicolumn{1}{c|}{4.248} &
   \multicolumn{1}{c|}{0.2310}
     \\ \hline
  \textbf{\begin{tabular}[c]{@{}c@{}}\ft only with Phase-2\end{tabular}} &
  \multicolumn{1}{c|}{398.0} &
  \multicolumn{1}{c|}{4.334} &
  \multicolumn{1}{c|}{0.4664}
    \\ \hline
  \textbf{\begin{tabular}[c]{@{}c@{}}\ft with both Phases \end{tabular}} &
  \multicolumn{1}{c|}{\cellcolor{blue!25} \textbf{366.0}} &
 \multicolumn{1}{c|}{\cellcolor{blue!25} \textbf{4.403}} &
  \multicolumn{1}{c|}{\cellcolor{blue!25} \textbf{0.5591}}
   \\ \hline
\end{tabular}
\label{tab:ablation_study_main}
\end{table}

\vspace{-2mm}
\subsection{Ablation study of \method}
We conduct an ablation study to confirm the contribution of each fine-tuning phase of \ft. 
Table~\ref{tab:ablation_study_main} shows the ablation study result of \ft with \opt text generation model on the \hmf~\cite{zhou2019content} model and Amazon Book dataset. As expected, the second phase (the rank-boosting fine-tuning) has a higher impact on the ranking performance than the first phase (domain-adaptive fine-tuning).
Conducting both phases of \ft leads to the best ranking metrics with statistical significance (p-values $< 0.01$).

\subsection{Hyperparameter Analysis of \ft}
\label{sec:exp:hyperparameters}
\begin{figure}[t!]
	\centering
	\begin{subfigure}[t]{0.48\linewidth}
		\includegraphics[width=4.0cm]{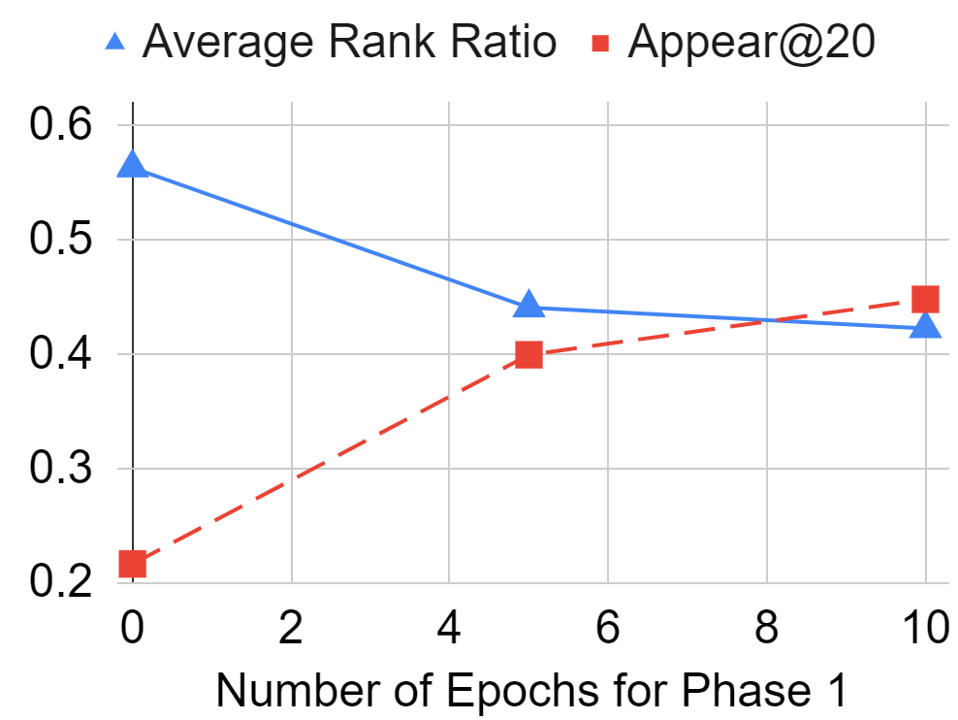}
		\label{fig:hyper_phase1_epochs}
	\end{subfigure}
	\begin{subfigure}[t]{0.48\linewidth}
		\includegraphics[width=4.0cm]{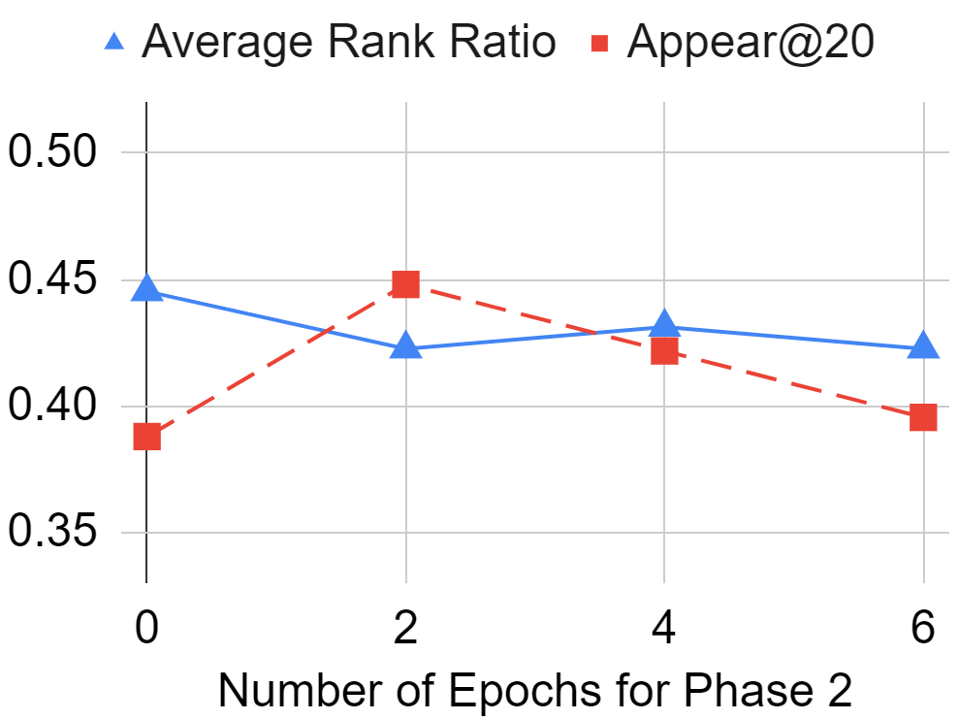}
		\label{fig:hyper_phase2_epochs}
	\end{subfigure}
	\begin{subfigure}[t]{0.48\linewidth}
		\includegraphics[width=4.0cm]{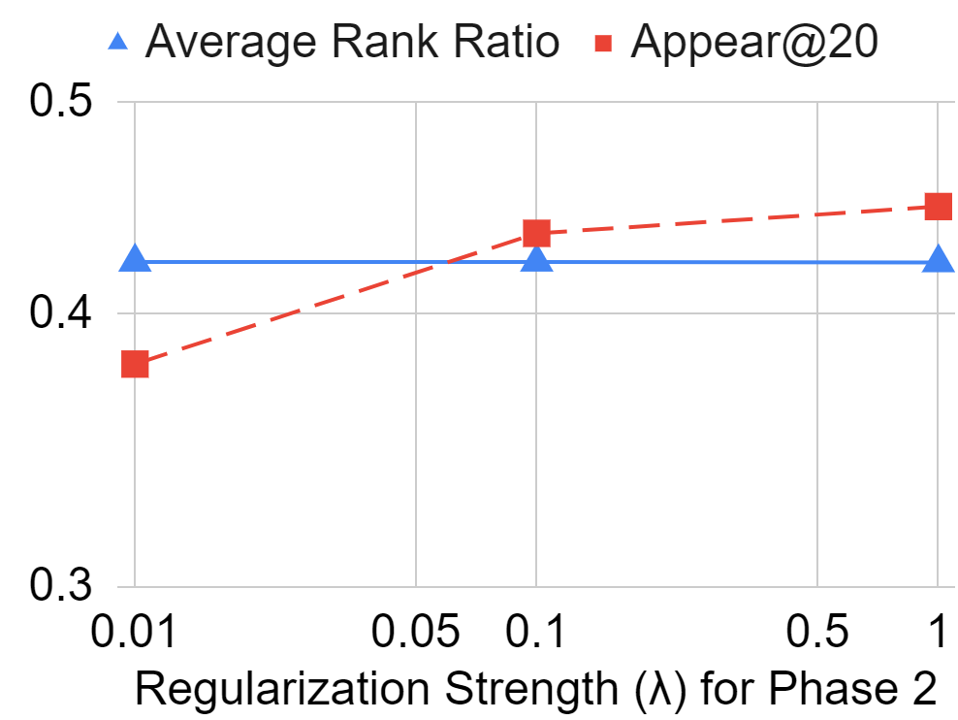}
		\label{fig:hyper_phase2_lambda}
	\end{subfigure}
	\begin{subfigure}[t]{0.48\linewidth}
		\includegraphics[width=4.0cm]{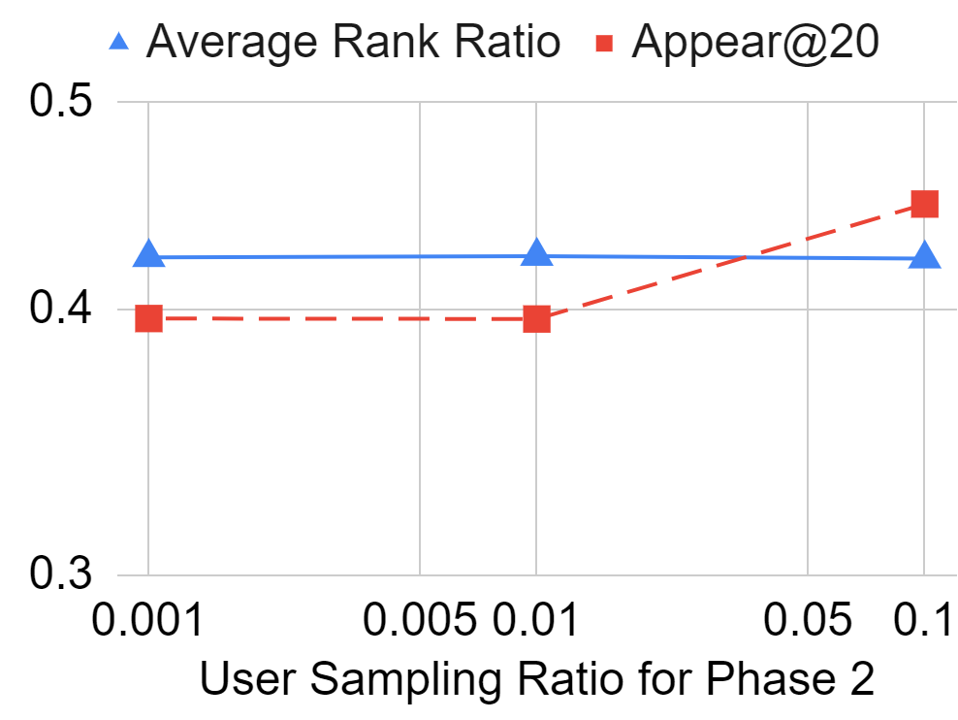}
		\label{fig:hyper_phase2_sampling}
	\end{subfigure}
	\caption{Hyperparameter sensitivity of \ft on \hmf model and Amazon Book dataset as per two ranking metrics. 
	}
	\label{fig:hyperparameter}
\end{figure}

Hyperparameters of \ft are chosen by the following. 
The learning rate is set to $10^{-5}$ among 
$[10^{-4},5\cdot10^{-5}, 10^{-5}]$, and the fine-tuning batch size is set to $16$ among 
$[16,32,64]$.
The numbers of Phase 1 and Phase 2 epochs are set to 10 (among 
$[2,5,10]$) and 2 (among 
$[2,5,10]$), respectively.
We randomly sample 1\% of the total items as target items. The size of sampled users and items $|\mathcal{U'}|$ and $|\mathcal{G'}|$ in Eq.~\eqref{eq:promotion_loss} are set to $0.1|\mathcal{U}|$ (among 
$[0.1|\mathcal{U}|, 0.5|\mathcal{U}|, 1.0|\mathcal{U}|]$) and $16$ (among 
$[16,32,64]$), respectively.
$\lambda$ in  Eq.~\eqref{eq:promotion_loss} is set to $1.0$ among 
$[0.01,0.1,1.0]$. 
$\lambda_1$ in  Eq.~\eqref{eq:promotion_loss} is set to $0.01$ among 
$[0.01, 0.1, 1.0]$.
Figure~\ref{fig:hyperparameter} presents the sensitivity analysis results of \ft with \pointer~\cite{zhang2020pointer} on the \hmf model and Amazon Book dataset as per two ranking metrics average rank ratio (=average predicted rank / number of candidates) and Appear@20. 
First, sufficient fine-tuning epochs in Phase 1 are required to absorb the linguistic background of the dataset; however, a small number of fine-tuning epochs is preferred in Phase 2 as too much fine-tuning after Phase 1 can lead to the overfitting of the text generation model (e.g., Appear@20 decreases after epoch 2). Regarding the rank promotion regularization, the performance of \ft is relatively stable to the regularization strength $\lambda$, $\lambda_1$, or sampling ratio of users and target items. Practically, these hyperparameters should not be too small or large to ensure effective promotion as well as accurate and fluent item descriptions.
We observe similar trends when we use other datasets and recommenders, and \ft with \opt~\cite{zhang2022opt}.

%% file: 070conclusion.tex
We proposed a text rewriting attack \method to promote the ranks of target items in text-aware recommender systems. With our two-phase fine-tuning and in-context learning approaches, \method can generate ranking-optimized item descriptions while maintaining the semantic meaning and fluency of the text. 
\method has several limitations such as (1) text rewriting quality can be lowered for rank promotion, (2) calculating the rank promotion loss can be costly for a large-scale dataset, and (3) \method can be sub-optimal for multimodal recommenders (e.g., using text+image).
Future work of \method includes (1) extending \ft with a text quality constraint and \icl with larger language models such as GPT-4 or LLama-3, (2) developing a defense mechanism against our text rewriting attack. Conducting \textbf{adversarial training} with rewritten text obtained by \method can be a promising \textbf{defense mechanism}, and (3) investigating other text and image manipulation attacks.

%% file: 080appendix.tex
\noindent \textbf{Surrogate Model Training for Black-box Attack.}
The first step is generating new surrogate training data for the surrogate model using fake user profiles. 
Similar to existing literature~\cite{yue2021black, zhang2021reverse}, we assume each fake user $z \in \mathcal{Z}$ can have at most $\mathcal{N}$ (e.g., 50) interactions or API calls to $\mathcal{M}$, and the total number of fake users is limited to $|\mathcal{Z}|$ (e.g., 5000). 
The surrogate data generation method depends on the type of $\mathcal{M}$ (e.g., sequential or collaborative filtering-based). Hence, we assume \textbf{we are aware of the type of $\mathcal{M}$ in advance.} 

If $\mathcal{M}$ is a sequential model, we use the Autoregressive Data Generation method, proposed by \cite{yue2021black}. It first initializes each profile with a random item and feeds the current profiles to the recommender system and obtains the top-K items for each profile. Next, it randomly samples one item from the top-K and appends the sampled item to each profile. It repeats the above process until $\mathcal{N}$ interactions are generated for each user. 
The autoregressive generation makes the new data resemble the original data distribution~\cite{yue2021black}.
If $\mathcal{M}$ is a conventional model, we use the Random Injection Collection method, proposed by \cite{zhang2021reverse}. 
First, all fake users randomly interact with $\mathcal{N}$ items. Next, for each fake user $z$, we randomly sample $\mathcal{N}$ items as queries, feed the current profile, and query items to the black-box recommender. We augment the $\mathcal{N}$ query items and their returned prediction scores for each user $z$ and construct the training data. 
The prediction scores of diverse query items are crucial indicators of the original training data distribution.

After generating the fake data, we use it to train a surrogate model $\mathcal{M'}$.  
We use \unisrec~\cite{hou2022unisrec} as $\mathcal{M'}$ if $\mathcal{M}$ is sequential; otherwise, we adopt neural collaborative filtering~\cite{he2017neural} as $\mathcal{M'}$. 
Following existing literature~\cite{tang2020revisiting, zhang2021data, yue2021black,zhang2021reverse}, we feed the surrogate training data to the surrogate model $\mathcal{M'}$ for training, which encourages $\mathcal{M'}$ to function like the black-box recommender $\mathcal{M}$. 
The loss function of training $\mathcal{M'}$ is given as follows.

$\mathcal{L}_{surrogate} = \sum_{\forall (u,i) \in \mathcal{X'}_{tr}} \{\Theta(u,i; \mathcal{M})-\Theta(u,i; \mathcal{M'})\}^2$,
while $\mathcal{X'}_{tr}$ indicates the surrogate training data, $\Theta(u,i; \mathcal{M})$ and $\Theta(u,i; \mathcal{M'})$ are scoring functions of  $\mathcal{M}$ and $\mathcal{M'}$, respectively. 